\documentclass[prd,aps,eqsecnum,tightenlines,nofootinbib]{revtex4}
\input epsf
\usepackage{graphicx}

\newcommand{\beq}{\begin{equation}}
\newcommand{\beqa}{\begin{eqnarray}}
\newcommand{\eeq}{\end{equation}}
\newcommand{\eeqa}{\end{eqnarray}}

\newcommand{\lsim}{\lesssim}
\newcommand{\gsim}{\gtrsim}
\newcommand{\psim}{\mbox{\raisebox{-1.0ex}{$~\stackrel{\textstyle \propto}
{\textstyle \sim}~$ }}}
\newcommand{\vect}[1]{\mbox{\boldmath${#1}$}}
\newcommand{\lmk}{\left(}
\newcommand{\rmk}{\right)}
\newcommand{\lnk}{\left\{ }

\newcommand{\rnk}{\right\} }
\newcommand{\lkk}{\left[}
\newcommand{\rkk}{\right]}
\newcommand{\lla}{\left\langle}
\newcommand{\p}{\partial}
\newcommand{\rra}{\right\rangle}
\newcommand{\so}{M_\odot}
\newcommand{\mch}{{\cal M}}
\newcommand{\vex}{{\vect x}}
\newcommand{\vel}{{\vect \Omega}_l}
\newcommand{\ven}{{\vect \Omega}_s}
\newcommand{\vep}{{\vect p}}
\newcommand{\veq}{{\vect q}}

\begin{document}
\title{Strong Gravitational Lensing and Localization of
Merging Massive Black Hole Binaries with LISA}
\author{Naoki Seto
}
\affiliation{Theoretical Astrophysics, MC 130-33, California Institute of Technology, Pasadena,
CA 91125
}

\begin{abstract}
We study how the angular resolution of LISA for merging massive black-hole
 binaries  would be
improved if we  observe multiple gravitational wave ``images''  due
 to strong gravitational 
lensing. The correlation between fitting parameters is reduced
 by the additional information of the second image which significantly
 reduces the error box 
 on the sky. This improvement would  be very
 helpful for identifying the host galaxy of a binary.
The angular resolution expected with multiple detectors is also discussed.
\end{abstract}
\pacs{PACS number(s): 95.55.Ym 04.80.Nn, 98.62.Sb }
\maketitle


\section{Introduction}

Black holes are a very fascinating prediction of general
relativity. Recent observations indicate that massive black holes (MBHs)
 exist at the centers of many galaxies and their masses have a strong
correlation with the properties of their host galaxies
\cite{Merritt:2001zg}. Binary MBHs would be 
formed 
by the merger of galaxies that harbor MBHs \cite{begelman}. 
The coalescence of such a binary is the most energetic event in the
Universe. It finally releases enormous energy  (several orders
smaller than 
 $c^5/G\sim 
 4\times 10^{59}$erg/sec) over a characteristic time scale $GM/c^3\sim
5(M/10^6\so)$sec ($M$: mass of the system). The Laser Interferometer
Space Antenna (LISA) can observe  gravitational waves from MBH
binaries with high signal to noise ratio (SNR) \cite{lisa}. In the case
of equal 
mass binaries the expected SNR at a given distance is  maximum
for redshifted masses $\sim 10^6\so$ and is $\gsim 10^3$
even at cosmological distances \cite{Cutler:1998ta}. By fitting the
gravitational waves 
from a MBH binary  with templates
prepared in advance  we can make a stringent test of general relativity
and obtain various important data about the black holes ({\it e.g.}
mass, spin). Therefore gravitational waves from MBH binaries are, by
themselves, a very 
interesting signal for relativistic astrophysics \cite{lisa,Cutler:2002me}.

If we can identify  their
host galaxies, it  would also have a 
significant impact in astrophysics and cosmology. By detailed follow-up
observations 
using electro-magnetic waves at various frequencies we might
discover absorbing properties of the galaxies or obtain fundamental
clues which will elucidate the evolution of MBHs. 
 In principle, the angular direction of coalescing MBH binaries could be
estimated by analyzing data from LISA.
Due to annual revolution and rotation of LISA the signal from a MBH binary
shows modulation that depends strongly on the ski position of the binary.
For equal mass  MBH binaries at 
$z\sim 1$ the angular resolution of LISA would be typically $\sim
10^{-4}$sr  for redshifted masses $\sim 10^6\so$
\cite{Cutler:1998ta}. The resolution 
becomes worse for high redshift MBH binaries.  The number of
bright galaxies is $\sim 10^4$ in one square degree ($\sim
3\times 10^{-4}$sr). Therefore the angular resolution of LISA is not sufficient
to fully specify the host galaxy of a cosmological MBH binary
\cite{Cutler:1998ta}.  Thus an 
improvement in the angular resolution for a MBH binary would be crucially 
helpful for selection of  candidates for its host galaxy using
electro-magnetic telescopes. Possible mechanisms providing such an
 improvement are worth exploring.

In this context it has been  discussed that higher gravitational wave harmonics of
 \cite{Moore:1999zw},  precession of the orbital plane in a
rapidly spinning 
system \cite{Vecchio:2003tn}, or effects caused by LISA's finite
arm-length of 
 \cite{Seto:2002uj} could become 
important in some situations. In this paper we show that the angular
resolution for a MBH binary could be significantly improved, if we
 observed multiple ``images'' of the gravitational waves due to 
strong
gravitational lensing. Our point is not the amplification of the signal
but observation of the same source  at different epochs.
The intrinsic binary parameters such as the chirp mass can be determined
very accurately from gravitational waves, so it would not be a hard
task to identify two lensed signals from the same  merging MBH binary.

If we assume that  MBH binaries  coalesce shortly after the merger of
their 
host galaxies  less than the
Hubble time, the hierarchical model of
structure formation suggests that the
coalescence rate of MBH binaries could be as high as 
$\sim100{\rm yr^{-1}}$, and its
distribution 
might be much higher at high redshift
than  low redshift $z\lsim 1$ \cite{Haehnelt:wt}. The probability $\tau$
of  strong 
gravitational lensing  also depends strongly on the redshift of the
source. At $z\sim 1$  the probability is  only $\tau\lsim
10^{-3}$, but  
it could reach several percent  at $z\sim 5$ \cite{Holz:1998uw}. Thus we
might 
obtain multiple images (more precisely the chirping gravitational
waves) of distant MBH 
binaries. In this paper it is shown that the  additional images caused by
lensing with time delay $\Delta T\gsim 0.1$yr  would greatly decrease the
correlations between fitting parameters of the templates, and the
angular resolution of a MBH binary could be dramatically improved.

This paper is organized as follows. In Sec II we briefly discuss signal
analysis for a single image (IIA) and for multiple images (IIB). Various
numerical results are presented in Sec III. We also describe the angular
resolution obtained with multiple detectors that are widely
separated (IIIC). Our study is summarized in Sec IV.

\section{Analysis}

\subsection{Observation of a chirping gravitational wave}
The three spacecrafts of
LISA maintain a nearly 
equilateral triangle configuration  and move annually around the Sun
with the 
orientation of the triangle  changing. As discussed below, the direction of a
gravitational wave source 
can be estimated from 
the wave signal which is  affected by (i) the Doppler effect due to the revolution of the
detectors around the Sun and (ii) the amplitude modulation due to the rotation of the
triangle and the increase of the wave frequency (chirp signal) \cite{lisa,Cutler:1998ta}.
For the response of LISA to gravitational waves we  follow
\cite{prince} for the time-delay interferometry (TDI) analysis that
cancels the laser frequency fluctuations in the data streams
\cite{Armstrong:uh}.

We label three spacecrafts as 1, 2, and 3 and use notation $L_i$
 ($i=1,2,3$) to denote the length of the arm opposite  spacecraft $i$. 
 Information about the  binaries is extracted  using  three
data streams $A(t), E(t)$ and $T(t)$ that form an orthogonal basis for the
 laser-noise canceling combinations.
These three data streams are constructed from six (one-way) relative
frequency 
 fluctuations $y_{ij}(t)$ ($i,j=1,2,3, i\neq j$) at different times
 (see  \cite{prince} for explicit expressions of $A,E,T$ in terms of
 $y_{ij}(t)$). The signal $y_{ij}$  is measured at spacecraft $j$ and
 transmitted from spacecraft $k(\ne i,\ne j)$ along the arm $L_i$.

Our analysis is similar to \cite{Cutler:1998ta} (see also \cite{Vecchio,Hughes:2001ya,Holz}).
However, as in \cite{Seto:2002uj} we do not use  the long 
wave approximation which is valid only  for incoming gravitational waves 
larger than the 
arm-length of LISA $L=5.0\times 10^6$km, corresponding to $f=c/2\pi
L\sim 0.01$Hz. 
The study in \cite{Seto:2002uj}  is somewhat incomplete, as the TDI analysis was not
included properly. This point is also  remedied in this paper.
For simplicity we assume that the arm-lengths $L_i$ are equal.

First we briefly discuss the  gravitational  waveform from a MBH
binary with a circular orbit. In this paper we only consider the in-spiral
waveforms. For larger MBH binaries the contribution of the merger or the
ring-down waveforms could
be important for SNR \cite{Hughes:2001ya}.  We use the stationary phase
approximation and 
the restricted post-Newtonian approach with 1.5-PN phase \cite{Cutler:tc} that is given in
Fourier space as follows
\beq
\Psi(f)=2\pi f t_c-\phi_c-\frac\pi4+\frac34(8\pi G c^{-3} \mch_z
f)^{-5/3} [ 1+\frac{20}9 \lmk \frac{943}{336}+\frac{11\mu_z}{4M_z}\rmk
x+(4\beta-16\pi) x^{3/2}  ], \label{1.5pn}
\eeq
where $t_c$ and $\phi_c$ are integration constants, and the former contains
information about the coalescence time measured at the Sun not at the
detector. The difference between the coalescence times is  included in 
eq.(\ref{fnc}).  The parameter $\beta$ is the 
spin-orbit coupling coefficient (we put its true value $\beta=0$), and $\mu_z$, $M_z$  $\mch_z$ are the reduced
mass, the total mass and  the chirp mass respectively. All of the mass
parameters are multiplied by the factor $(1+z)$ ($z$: redshift of  the
binary) with the suffix $z$. The total mass $M_z$ is given by other two
masses as $M_z=
\mch_z^{5/2} \mu_z^{-3/2}$.
The post-Newtonian expansion parameter  $x=O(v^2/c^2)$ is defined as
$x\equiv \lnk 
G \pi c^{-3}
M_z f 
\rnk^{2/3} $. 

For simplicity we first consider gravitational waves propagating in a
homogeneous background.  The effects of lensing will be discussed later.
The incoming gravitational wave from a binary is decomposed into  plus
and cross 
polarization modes. It is convenient to use the principle axes
$(\vep,\veq)$ defined by the 
direction $\ven$ and orientation $\vel$ of the binary as
$\vep={\ven\times\vel}/{|\ven\times\vel|}$ and $\veq=-\ven\times \vep$.
The plus mode has the polarization tensor $e^+_{ab}=p_ap_b-q_aq_b$ and
the cross mode has $e^\times_{ab}=p_aq_b+q_ap_b$.
The relative frequency shift, for example, $y_{31}(t)$,  due to these two
modes is 
expressed as 
\beq
y_{31}(t)=
\frac12  (\pi f)^{2/3}\lmk\cos2\psi_{12} A_+ +i\sin 2\psi_{12} A_\times \rmk
 (1-\cos
\theta_{12})  \lkk U(t,1)- U(t-\tau,2) \rkk, \label{phase}
\eeq
where 
$\tau$ is the time $\tau=L/c\sim 17$sec
and the function 
\beq
U(t,j)=\exp[-2\pi i f (t+\vex_j\cdot\ven/c)]\label{fnc}
\eeq
describes the phase of the incoming wave at  the detector $j$ (with
position $\vex_j$ measured from the Sun) at
time $t$. In Eq.(\ref{phase}) we have  neglected the very small relative
motion of the detectors over the 
time scale $\tau$ and also  the time variation of the frequency over the
time scale $1{\rm AU}/c=500$sec.  The coefficients $A^+$ and $A^\times$
are given in terms of 
an amplitude $A$ 
 as 
$
A^+=A (1+\cos^2I)$ and $A^\times =2A\cos I$ in which  $\cos I\equiv \ven\cdot\vel$ is the cosine of the
inclination angle. In Eq.(\ref{phase})
 $\theta_{jk}$ is the angle between the direction of the binary
$\ven$ and the arm $\vex_j-\vex_k$,  and  the principle polarization angle $\psi_{jk}$ is given by
$
\tan \psi_{jk}=[(\vex_{j}-\vex_k)\cdot \veq]/[(\vex_{j}-\vex_k)\cdot\vep].
$ The  angles  $\theta_{jk}$ and $\psi_{jk}$ change with the motion of the
spacecrafts.

The Doppler
effects are included in the factor $U(t,j)$.
The TDI data streams at time  $t$ are constructed from linear
combinations of 
$y_{ij}$ at different times, {\it e.g.} $t$, $t-\tau$ or $t-2\tau$. Due
to the finite arm-length,  the response to gravitational waves
shows a complicated behavior that depends strongly on the angular position
of the MBH binary.
The effect of  finite
arm-length on the 
estimation of the source direction could become important  around  $\sim
10^5+10^5\so$ in the case of equal mass binaries \cite{Seto:2002uj}.
In the long wave limit $f\tau\ll 1$ the above expression (\ref{phase})
is simplified.  The analysis by Cutler \cite{Cutler:1998ta} is
essentially in  this 
limit, and the amplitude modulation and phase modulation from Doppler
effects can be analyzed separately. As a comparison we also examine the
angular resolution of MBH binaries using this method to  describe
the  response of LISA
to gravitational waves.

The  intrinsic detector noise  for a
measurement of  the fluctuations $y_{ij}$ has two main  components that
are 
relevant for the combinations $A, E, T$. They are the proof mass noise
and the optical path noise. We use the spectra \cite{Armstrong:uh}
\beqa
S_y^{proof-mass}(f)&=&2.5\times 10^{-48}\lmk\frac{f}{1\rm Hz}
\rmk^{-2}{\rm Hz^{-1}},\label{pmn}\\
S_y^{optical-path}(f)&=&1.8\times 10^{-37}\lmk\frac{f}{1\rm Hz} \rmk^2{\rm
Hz^{-1}}. \label{opn}
\eeqa
It is a simple task to obtain the noise for the three combinations
$A,E,T$. 
But we note that the detector's noise  below 0.1mHz is somewhat
controversial (see {\it e.g.} \cite{Bender:uv,Hughes:2001ya}).  This frequency region strongly 
affects our
results for larger MBH binaries.

We also include the Galactic binary confusion noise \cite{Bender:hs}
 using the following 
simple approximation 
\beq
S^{Gc}_{h}(f)  =\cases{  
4\times 10^{-37} \lmk\frac{f}{1\rm mHz} \rmk^{-7/3}  {\rm
Hz^{-1}} &  $(f\le 10^{-2.7}{\rm
Hz}) $ \cr
0 &  $(f> 10^{-2.7}{\rm Hz})$ \cr}\label{bcn}
\eeq 
given for  one year
observation. 
This confusion noise is about 10 times larger (for $h$) than the
detector noise at 
1mHz (see figure 5 of \cite{prince}). For simplicity we neglect the anisotropies of the Galactic
confusion noise in this paper. 

We integrate each chirping gravitational wave for 1yr before 
coalescence up to the cut-off frequency $f_{max}$ when the binary
separation becomes $r=6GM_z/c^2(1+z)$, corresponding to a frequency
\beq
f_{max}=0.04 \lmk \frac{M_z}{10^5\so}\rmk^{-1} {\rm Hz}.\label{fmax}
\eeq 
In contrast the wave frequency $f$ observed 
at $T$yr  before the coalescence time $t_c$ is   given approximately
by
\beq
f=1.9\times 10^{-4} \lmk\frac{\mch_z}{0.87\times 10^5 \so}  \rmk^{-5/8}
\lmk \frac{T}{1{\rm yr}} \rmk^{-3/8} {\rm Hz}.\label{ftrel}
\eeq
The signal to noise ratio SNR of a  detection is determined by the
amplitudes of 
the modes $A, E$ and $T$ and by  the noise  spectra $S_A(f), S_E(f)$ and
$S_T(f)$ which are constructed from the detector  noises (\ref{pmn})
(\ref{opn}) and the
confusion noise (\ref{bcn}). 
We calculate the signal to noise ratio of the in-spiral waves  using
\cite{thorne, Finn:1992wt} 
\beq
{\rm SNR^2}=4\sum_{B=A,E,T}\int_{f_1}^{f_{min}}\frac{B(f) B^*(f)}{S_B(f)}df, \label{snr}
\eeq
where $f_1$ is the initial frequency for signal  integration.
In the present study the number of  fitting parameters $\{\alpha_i\}$
for each chirping gravitational wave 
is 10. They are
$\{\alpha_i\}=\{\mch_z,\mu_z,\beta,t_c,\phi_c,A,\theta_s,\phi_s,\theta_l,\phi_l\}$.
Angular variables $(\theta_s,\phi_s)$ represent the direction of the
binary $\ven$ in a fixed  frame around the Sun, and  $(\theta_l,\phi_l)$
represent its 
orientation $\vel$. The
variance in the parameter estimation errors $\Delta \alpha_i$ are
evaluated using the Fisher information matrix $\Gamma_{ij}$ as $\lla
\Delta\alpha_i \Delta\alpha_j  \rra=\Gamma_{ij}^{-1}$ where
$\Gamma_{ij}$ is given as \cite{Finn:1992wt} 
\beq
\Gamma_{ij}=4\sum_{B=A,E,T}\int_{f_1}^{f_{min}}\frac{\p_i B(f)
\p_j B^*(f)}{S_B(f)}df.  \label{fisher}
\eeq
 Following Cutler \cite{Cutler:1998ta}
we use the notation $\Delta \Omega_s\equiv 2\pi \sin \theta_s\sqrt{\lla
\Delta\theta_s^2\rra \lla \Delta\phi_s^2\rra -\lla
\Delta\theta_s\Delta\phi_s\rra^2 }$
for the angular resolution of a binary on the sky.

Our results for the angular resolution $\Delta\Omega_s$ change only
slightly for masses $\gsim 10^4\so$ when we switch the integration time
from 1yr to 10 yr. We put the
MBH binaries  at  $z=3$ in a  universe with 
cosmological 
parameters $\Omega_0=0.3$, $\lambda_0=0.7$ (spatially flat universe) and
$H_0=71$km/sec/Mpc. In 
this cosmological background  the 
luminosity distance $d_L$  becomes $d_L=25$Gpc at $z=3$ and $d_L=6.5$Gpc
at $z=1$. 

\subsection{Analysis for multiple images by lensing }

We discuss  parameter fitting of  multiple images created by
strong gravitational lenses. As we show below the improvement in the
angular 
resolution 
$\Delta\Omega_s$ due to lensed multiple images could be very effective
for time delays 
$\Delta T\gsim 0.1$yr. On the other hand a
time delay larger than the operation period of LISA (nominally $\sim3$
yrs, 
but possibly $\sim$10 yrs) is
 irrelevant for our analysis. Thus our main targets are multiple images
with time delay $0.1{\rm yr}\lsim \Delta T \lsim 10 $yr. LISA has good sensitivity to
gravitational waves with frequency  $10^{-4}{\rm Hz}\lsim f \lsim 10^{-1}{\rm Hz}$
which is much higher than the inverse of the relevant time delay 
$(\Delta T)^{-1}\lsim 3\times
10^{-7}$Hz.  We note also that the latter frequency is smaller than $f$
given by
eq.(\ref{ftrel}) for $T=1$yr and $\mch_z\lsim 10^8\so$. Therefore  the
geometric optics approximation is valid in our analysis and
 the structure of phase is not changed by the lensing \cite{Nakamura:1997sw,Takahashi:2003ix}. 
We take the same parameters $\mch_z,\mu_z,\phi_c,\beta$ for the phases
eq.(\ref{1.5pn}) of  all the
multiple images from a source, 
but the coalescence time $t_c$ is, of course, different.
In some cases the substructure of  the lensing galaxy might complicate the
problem \cite{Mao:1997ek}.

 In the case
of a homogeneous 
and isotropic universe the wave amplitude $A$ is  given simply in terms
of the 
luminosity distance $d_L$ and the chirp mass $\mch_z$ by \cite{Schutz:gp}
\beq
A_0=\frac{2G^{5/3}\mch_z^{5/3}}{c^4d_L}.\label{a0}
\eeq
The above relation can be used to  determine $d_L$ from the observed
gravitational wave and also to specify the
redshift of the binary $z$ by inversion of the $d_L$-$z$ relation if the
cosmological parameters are known accurately \cite{Schutz:gp,Hughes:2001ya}. Note that in general the chirp mass
$\mch_z$ is determined more accurately than the amplitude $A$ using the
time 
evolution of the gravitational wave phase.

For signals  affected by
strong lensing we fit  different  values $A_i$  for the
amplitude  of each   image.  We cannot apply the above inversion to
estimate $z$, 
but this  is 
not a serious problem for identifying high-z MBH binaries, considering
the accumulated effects 
of weak 
lensing. At $z\sim3$ the rms fluctuations of the amplification by weak
lensing  could
become $\sim 0.2$ and hamper the   application of the above inversion
even for 
a single image \cite{Holz,Barber:2000pe}.

Here we discuss fitting of the angular variables
$(\theta_s,\phi_s,\theta_l,\phi_l)$. First  we evaluate the typical image
separation for a lensed source.  As a model we use the  singular
isothermal 
sphere approximation for the density profile of the lensing galaxy.
In  this   model   we have the
following explicit relation between the image
separation angle $\theta$ and the time delay $\Delta T$ \cite{lens}
\beq
\Delta T=\frac{\theta^2}{2c}\frac{D_{OL} D_{OS}}{D_{LS}}(1+z_L)y
=1.3\lmk\frac{\theta}{5^{''}}\rmk^2\lmk
\frac{D_{OL}}{{D_{LS}}}\rmk 
\lmk\frac{D_{OS}}{1.4{\rm Gpc}} \rmk \lmk \frac{1+z_L}{2}\rmk 
\lmk
\frac{y}{0.5}\rmk {\rm yr},\label{singular}
\eeq
where $D=d_L/(1+z)^2$ is the angular diameter distance ({\it e.g.}
$D_{OL}$: between the observer and the lens), $z_L$ is the redshift
of the lens, and $y$ is the normalized impact parameter with $y<1$
corresponding to the 
occurrence of multiple images in  the isothermal sphere
approximation. 
  The  image separation ($\lsim 5^{''}$) for the relevant
time delay
$\Delta T \sim 1$yr is
much smaller than 
the angular 
resolution of the source direction for cosmological MBH binaries,  which
will be shown later. This
means that
we can use the same fitting parameters $(\theta_s, \phi_s)$ for the
direction of the binary for each image.

The polarization tensor ${\bf e}_{ab}$ of a gravitational wave  is
parallel 
transported along a null geodesic  
\cite{gravitation}. It is easy to confirm that the  directions of the
tetrad 
vectors on  a lensed null 
geodesics  change only on the order of
the image separation, which is generally much smaller than the resolution
of the orientation of a cosmological MBH binary
$(\theta_l,\phi_l)$. This means that the
polarization properties
of the 
multiple  images cannot be distinguished observationally  and   
 the same fitting parameters
$(\theta_l,\phi_l)$ can also be used for the orientation of  each
image.  
 
Let us briefly summarize the number of fitting parameters.
For  $n$-multiple images from the same source we can take eight
common parameters 
$\{\mch_z,\mu_z,\beta,\phi_c,\theta_s,\phi_s,\theta_l,\phi_l\}$, but fit
two different ones $\{A_i,t_{ci}\}$ for each image ($i=1,...,n$). Thus the total number of
fitting parameters becomes $8+2n$.
In the following analysis we fix the true value of  each wave
amplitude $A$ (both un-lensed and lensed) by $A=A_0$. This would give
conservative results  
 for the magnitude of the parameter estimation errors in the case of
lensed 
 signals. With our 
prescription the SNR depends on the total number of images $n$ as
$SNR\psim 
n^{1/2}$.

\section{Angular resolutions for MBH binaries}
\subsection{Basic analysis}
We calculate the parameter estimation errors using  (i) the TDI
method described in Sec II and (ii) the method of Cutler \cite{Cutler:1998ta} which uses  the
long-wave 
approximation. The effective noise curves are somewhat different for these
two approaches. In this section various averaged quantities, such as the SNR and the angular resolution
$\Delta \Omega$, are evaluated by taking geometrical averages for 100 MBH
binaries at $z=3$ with  random directions $(\theta_s,\phi_s)$ and
orientations  $(\theta_l,\phi_l)$. For simplicity we only study equal
mass binaries. The time delay is fixed at $\Delta T=1/3$yr unless
otherwise stated. 

In Figure 1 the averaged the SNR is presented for redshifted  masses
$4\times10^3-4\times10^8\so$ (true masses $10^3-10^8\so$). Hereafter we
mainly use the redshifted mass $m_{z}=m_{1z}=m_{2z}$ to show the mass
dependence. 
When we decrease the mass from $m_z\sim10^8\so$, the SNR becomes a maximum around
$m_z\sim 10^6\so$. 
This is because the binary confusion noise disappears around $f\sim
10^{-2.7}$Hz in our models and this frequency corresponds to the
coalescence frequency $f_{max}$ of a  MBH binary with mass $m_z\sim
10^6\so$ as given by  eq.(\ref{fmax}). The SNR for two lensed  images
simply 
increases by a 
factor of $\sim 1.5\sim \sqrt2$ as  expected.

In Figure 2 we show the averaged angular resolution as a function of
redshifted mass $m_z$. For single images the averaged resolution $\Delta
\Omega_1$ shows weak dependence on the mass for the TDI method, and is
nearly constant at $\sim  2\times 10^{-3}$sr for MBH binaries with masses
between $m_z\sim 
10^5\so$ and
$10^8\so$ at $z=3$.  This is a remarkable contrast to  Figure 1 for
SNR which shows a steep rise around $m_z\sim10^5-10^6\so$.
Using the simple  method of Cutler we find that the sky positions of a MBH
binaries  
with  $m_z\gsim 10^5\so$ is mainly estimated from the amplitude
modulation due to rotation of the detectors. The 
Doppler phase modulation has a  contribution for $m_z\lsim 10^5\so$.

When a second image is added by the time delay of lensing, the situation
changes 
drastically. The angular resolution $\Delta \Omega_2$ obtained from two
images  with
$m_z\sim 4\times 10^5\so$ is improved by  more than two orders of
magnitude, compared to a  single image $\Delta \Omega_1$. The ratio
$\Delta \Omega_1/\Delta \Omega_2$ 
decreases for  both smaller and larger masses. It becomes
$\sim 10$ at $m_z\sim  10^4\so$ and $\sim 5$ at $m_z\sim  10^8\so$. This mass
dependence is discussed later. We also found that the parameter
estimation  errors for the
intrinsic binary parameters such as the chirp mass would be changed by
only a
factor of  $2$ or so by the second image.

In Figure 3 we show  histograms of the angular resolutions $\Delta
\Omega_1$ and $\Delta \Omega_2$ in our 100 realizations of MBH
binaries with redshifted masses $ m_z=4\times 10^3\so,  4\times 10^5\so$ and
$ 4\times 10^7\so$. The impact of  lensing is also apparent in these
figures.

So far we have fixed the time delay at $\Delta T=1/3$yr. In Figure 4 we
present the ratio $\Delta \Omega_1/\Delta \Omega_2$ as a function of the
time delay $\Delta T$. Due to the periodicity in the configuration of
LISA,  the results obtained  for $\Delta T$yr  are the  same as those
for  $\Delta
T+N$yr with $N$ an 
integer.  As our results are given in the form of a  ratio $\Delta
\Omega_1/\Delta \Omega_2$, they do not depend on the distance or
redshift to  the source  as long as we use the redshifted masses.
The factor $\Delta \Omega_1/\Delta \Omega_2$ depends  only weakly on the
time delay for  $\Delta T\gsim
0.1$yr. 

\subsection{Third image}

We have also calculated the angular resolution $\Delta \Omega_3$ in the
case that we could observe a  total of three images. We set the two time
delays relative to  the first image as  $\Delta T=1/3$ yr and $\Delta T=2/3$ yr,
and fixed the amplitudes of the three images by $A_0$ given in
eq.(\ref{a0}). The 
ratios  $\Delta \Omega_1/\Delta \Omega_3$
 become  $\sim 40$   ($m_z=4\times  10^3\so$),  $\sim 800$   ($4\times
10^5\so$) and  $\sim 22$   ($4\times
10^7\so$).  For the first two images analyzed in the previous
subsection the ratios  $\Delta \Omega_1/\Delta \Omega_2$ are $\sim 13$   ($4\times  10^3\so$),  $\sim 300$
($4\times 
10^5\so$) and  $\sim 9$   ($4\times
10^7\so$). Therefore the effect of a third image is not so drastic as
that of the second one.

\subsection{Multiple detectors}
We have shown that the observation of multiple images due to lensing could
significantly 
improve the angular resolution $\Delta \Omega$. But we cannot bet on 
this passive effect for most merging MBH binaries, considering the
probability. In an usual situation
we will 
only detect a single image. Using multiple detectors is one
positive method for improving the angular resolution. In this case we can take the
same fitting parameters also for the amplitude $A$ and the coalescence
time 
$t_c$ (at the Sun). 
The former is important as the amplitude $A$ and the angular variables
will correlate strongly (see eq.(\ref{phase})).
The latter means that we can take advantage of the time delay between
the two 
detectors 
which is closely related to  the direction to the  binary.

As a simple extension of our analysis for lensing,  we calculate the
angular resolution $\Delta \Omega_{II}$ with two detectors that have the
same specification as LISA. The two
detectors are on the Earth orbit around the Sun as for LISA, but we set
the   angle between them at  $2\pi/3$ corresponding to 1/3yr of  orbital
time. Thus their 
distance is fixed at 1AU$\times2\times \sin(\pi/3)=$1.73AU, and their
orientations are always different from each other.  For simplicity we
assume that the noises 
in the data streams measured by the two detectors are independent. This
would be a valid approximation 
for the detector noise, but might not be for the binary confusion noise.
We evaluate the Fisher matrix using an extension of eq.(2.10) with six
data streams; $(A,E,T)$ modes both for the two detectors.

The ratio of angular resolutions for a single LISA $\Delta \Omega_1$ and
for two LISAs $\Delta \Omega_{II}$ becomes $\Delta \Omega_1/\Delta \Omega_{II}\sim 10^2$ ($4\times
10^3\so$),   
$\sim 6\times 10^3$ ($4\times 10^5\so$) and $\sim 30$ ($4\times
10^7\so$).  Therefore even for binaries at $z=3$ the angular position of
a merging MBH 
binary can be determined with error box $\sim  10^{-6}$sr, namely 
(several arc-minutes)${}^2$. This area on the celestial sphere is close to
the 
resolution obtained from
Gamma-Ray-Bursts for  determining their host galaxies. 
When  another detector (a total of  three) is added 
 with positions characterized
by 1/3yr and 2/3yr, the ratio $\Delta \Omega_{1}/\Delta\Omega_{III}$
becomes $\sim 400$, $\sim 4\times 10^4$ and $\sim 60$ respectively.

\subsection{Effective time}
As shown above, the angular resolution $\Delta \Omega$ could be
 improved significantly for $m_z=10^5\sim10^6\so$. Here we
analyze this mass dependence. Roughly speaking, the SNR characterizes the
magnitude of the Fisher  matrix $\Gamma_{ij}$ as we can understand from
eqs.(\ref{snr}) and (\ref{fisher}). When the correlation (degeneracy) between the fitting
parameters is large, the parameter estimation errors such as $\Delta
\Omega$ also become large. This correlation  depends
strongly on the configuration of LISA around the epoch when the SNR
accumulates.  In this situation some independent information {\it e.g.}
added by the 
second image,  has an important role 
in reducing the correlation and improving the angular
resolution $\Delta \Omega$. This fact can be anticipated by comparing
Figures 1 and 2.

The angular directions of  MBH binaries are mainly estimated from the
amplitude modulation and the Doppler phase modulation caused by the
revolution and 
rotation of LISA. A long duration observation is crucial for determining
them well. Therefore the correlation between parameters including 
the direction would be  related to the effective observational
period, and a binary  with a shorter observational period would  have a larger
ratio $\Delta \Omega_1/\Delta \Omega_2$. 
Let us define a time $t_{eff}$ by weighting the time before coalescence $t_c-t$
by the amplitude of the wave $h(f)$ and the noise curve $S_h(f)$ as
follows
\beq
t_{eff}^2\equiv \frac{\lmk4 \int\frac{h(f) h^*(f) (t-t_c)^2}{S_h(f)}df
\rmk}{\lmk4 \int\frac{h(f) h^*(f)}{S_h(f)}df
\rmk} .\label{eft}
\eeq
Here we formally wrote down the above expression (see eq.(\ref{snr})). The
denominator 
is nothing but the square of the SNR. As  the weight factor
$|h(f)|/S_h(f)^{1/2}$ does not have a very  spiky structure,  we can
regard $t_{eff}$ as an effective observational period of the binaries.  For
monochromatic 
sources we  have $t_{eff}=1/\sqrt{3}$yr. In figure 5 we show the time
$t_{eff}$ (solid curve) and the ratio of the angular  resolution $\Delta
\Omega_2/\Delta\Omega_1 $ (dashed curve). The overall shapes of the curves
are similar as expected. 

The mass dependence of $t_{eff}$ can be understood as follows. For 
larger mass binaries with $m_z \gsim 10^5\so$,  the SNR (denominator of
eq.(\ref{eft}))   comes 
mainly from the  wave emitted  close to  the final coalescence.  But these
waves have a  small contribution to  the numerator.  A binary with mass
$m_z\lsim 10^6\so$ has coalescence frequency larger than $10^{-2.7}$Hz,
which is critical for the Galactic binary
confusion noise.  Therefore the effective observational
time $t_{eff}$ decrease significantly for $m_z\lsim 10^6\so$.
The smaller mass ($m_z\lsim 10^4\so$) binaries   stay in
the sensitive frequency region for a longer time, and the effective time $t_{eff}$
increases.

\section{summary}
In this paper we have studied how the angular resolution of LISA would be
improved if we  observe multiple images due to strong gravitational
lensing. It is  found that the error box on the sky could be typically
100 times smaller  for redshifted mass $m_z=10^5\sim 10^6\so$. The
improvement in the angular resolution
depends strongly on the masses of the MBH binaries.  
This mass dependence can be roughly explained  by  using an  effective
observational time for the  binary.  

We have mainly discussed the effects of lensing on the matched filtering
analysis  of
gravitational waves, but lensing  would  also be of great advantage
in  searching for the host galaxy of a MBH binary using electro-magnetic
waves. 
Besides the fact  that the target itself  is lensed, additional information
obtained from the gravitational waves, such as the time delay or the 
ratio of
the amplification factors of the images, could be used to  specify the
host 
galaxy.

We have also calculated the angular resolution expected for multiple
detectors and found that the resolution could be improved by a factor of 
$\sim  6\times 10^3$ ($m_z\sim 4\times 10^5\so$) compared to  that from  a single
detector. 
If LISA could detect gravitational waves from  MBH binaries, very
exciting results might be obtained by launching  another detector.

\begin{acknowledgments}

The author is grateful to  J. Gair for carefully reading the
 manuscript. 
He also thanks
R. Takahashi for valuable discussions, and 
 S. Larson  and A. Cooray for helpful comments.  

\end{acknowledgments}


\begin{figure}
  \begin{center}
\epsfxsize=8.cm
\begin{minipage}{\epsfxsize} \epsffile{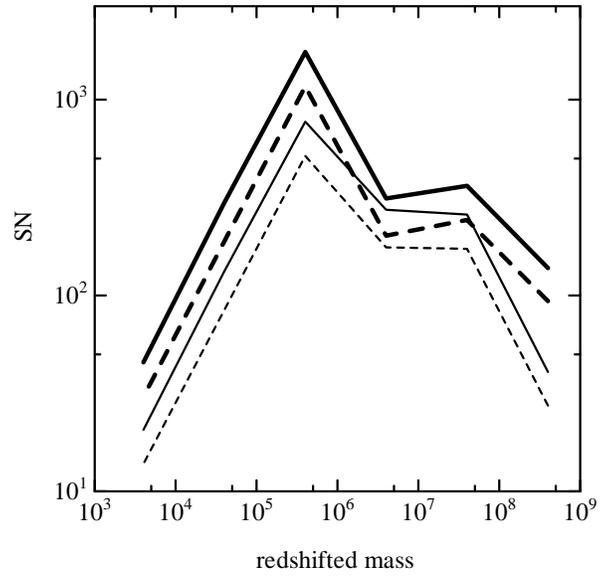} \end{minipage}
 \end{center}
  \caption{Averaged SNR for equal mass binaries at $z=3$. The thick
 curves are for the TDI method and the thin ones are  for the method of
 Cutler. The 
 results for the single (double) image are given by the dashed (solid) curves
 respectively. The horizontal axis represents the redshifted mass $m_z$.}
\label{f1}
\end{figure}

\begin{figure}
  \begin{center}
\epsfxsize=8.cm
\begin{minipage}{\epsfxsize} \epsffile{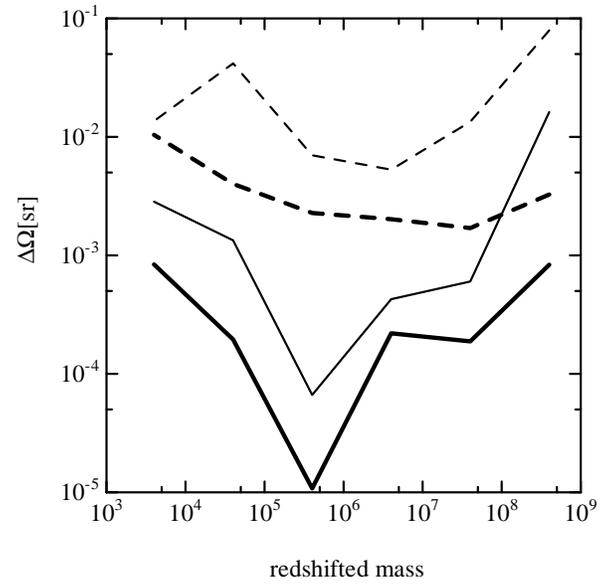} \end{minipage}
 \end{center}
  \caption{Averaged angular resolution. The identification of each
 curve is the same as in  figure 1.}
\label{f2}
\end{figure}

\begin{figure}
\begin{center}
\begin{tabular}{ccc}
(a)\includegraphics[scale=0.38]{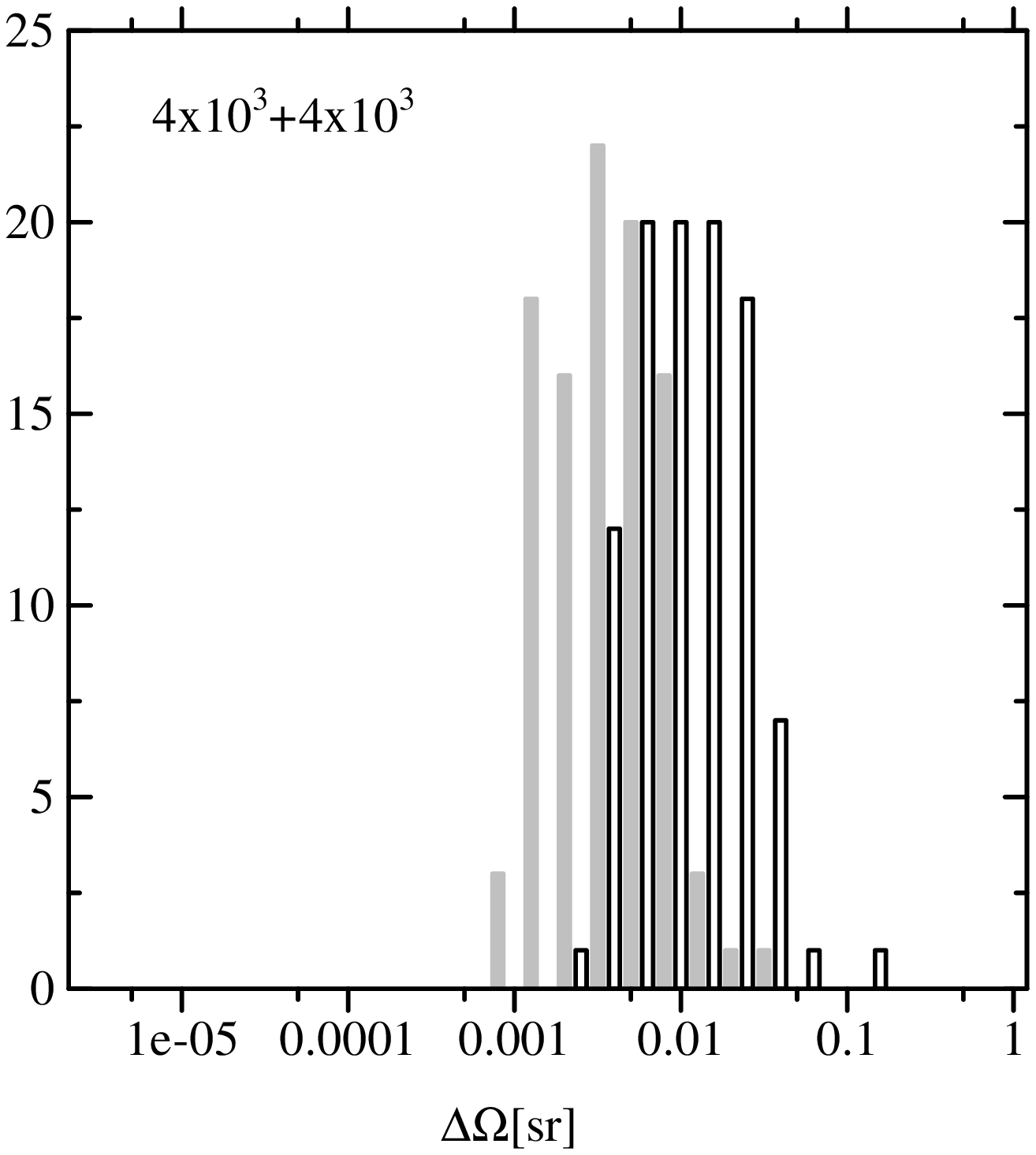} &
(b)\includegraphics[scale=0.38]{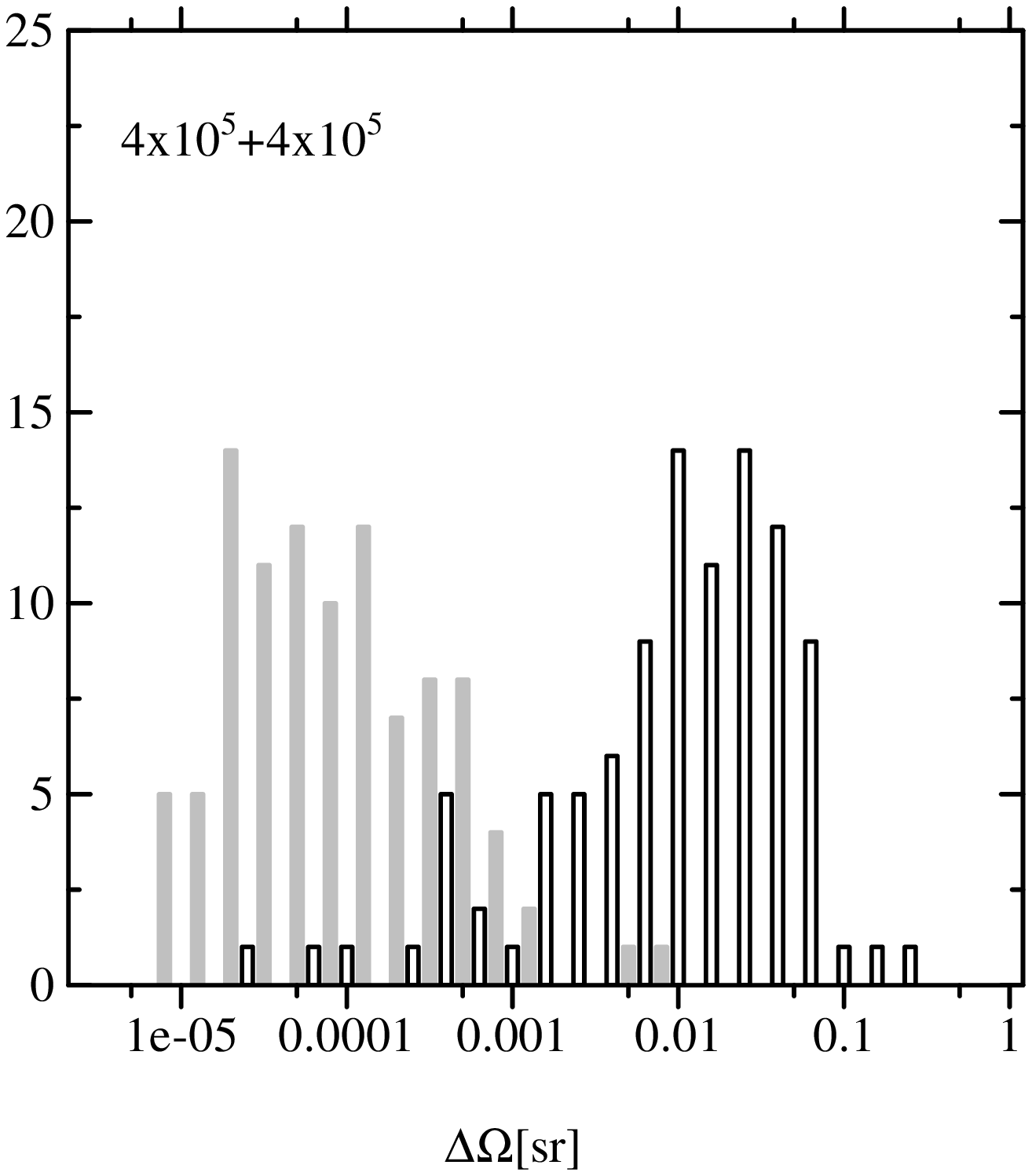} &
(c)\includegraphics[scale=0.38]{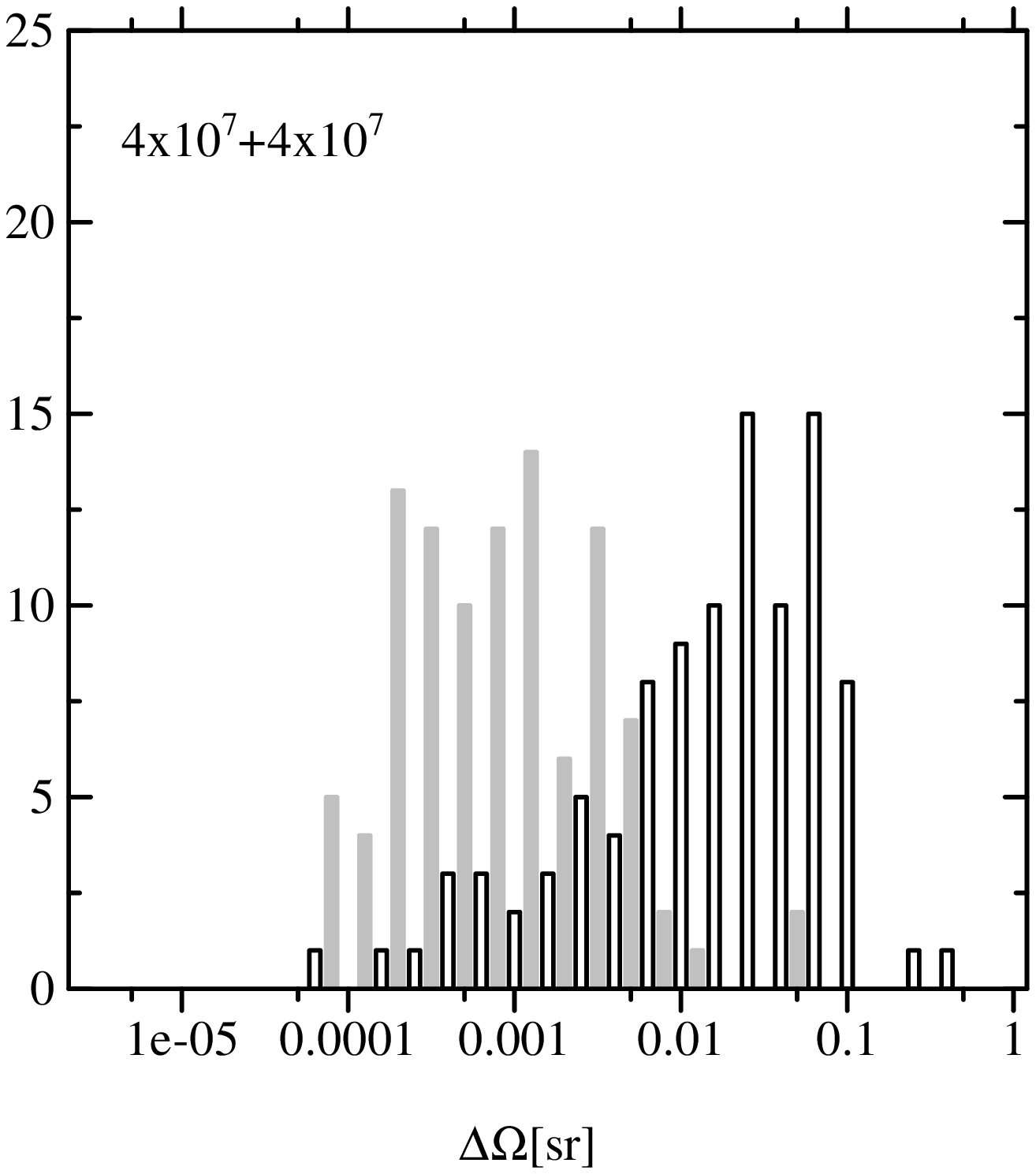} \\
\end{tabular}
\end{center}
\caption{Histograms of the angular resolution $\Delta \Omega_1$ (open
 bars) and $\Delta \Omega_2$ (gray bars). The mass $m_z$ is fixed at
  (a) $m_z=4\times 10^3$,  (b) $4\times 10^5$ and (c) $4\times 10^7$. The
 results for $\Delta \Omega_1$ are displaced by $10^{-0.1}$. The total number
 of binaries is 100 for each  mass. These results were obtained using
 the TDI method.}
\label{f3}
\end{figure}

\begin{figure}
\begin{center}
\includegraphics[scale=0.6]{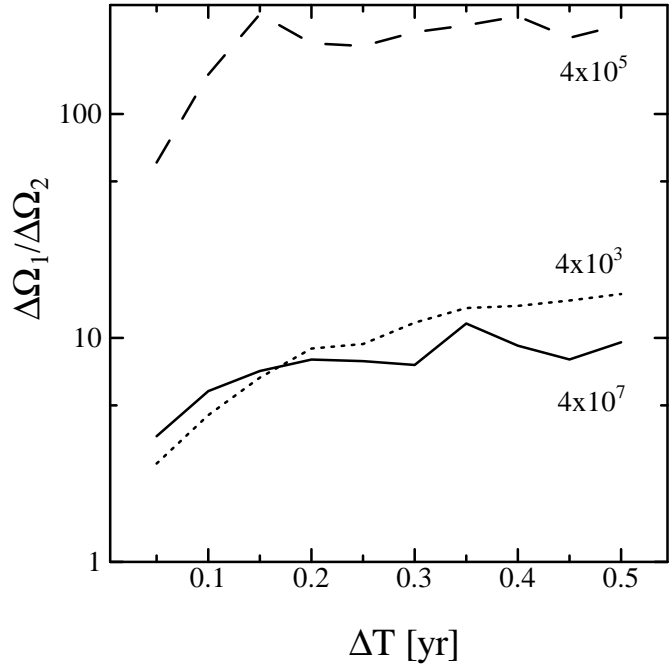} 
\end{center}
\caption{Averaged ratio $\Delta \Omega_1/\Delta \Omega_2$ as a function
 of the time delay $\Delta T$.  The three curves represent different
 masses. These results do not depend on the distance to  the binaries.
 The TDI method  was used again.}
\label{f4}
\end{figure}

\begin{figure}
\begin{center}
\includegraphics[scale=0.6]{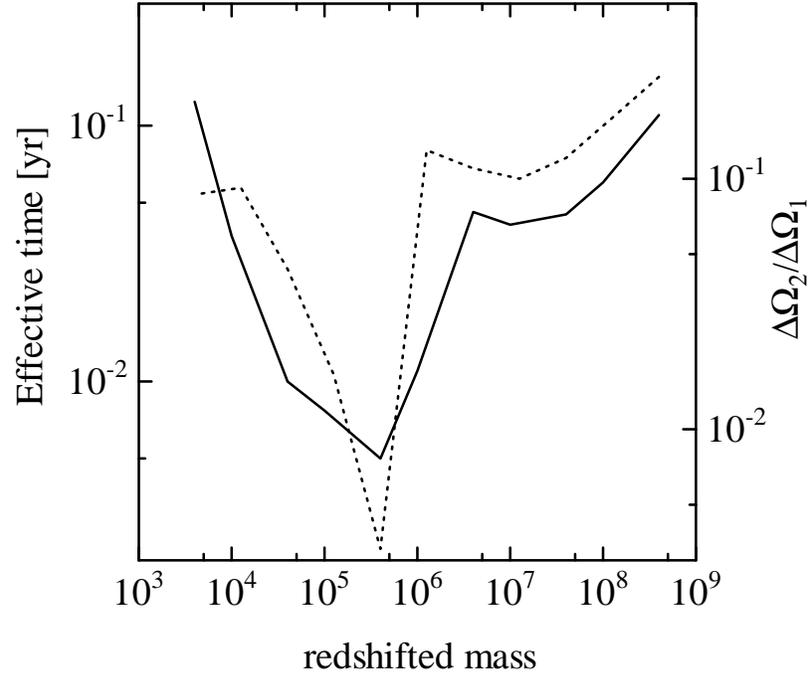} 
\end{center}
\caption{The effective time $t_{eff}$ (solid curve: left axis) and the ratio
 $(\Delta \Omega_1/\Delta \Omega_2)^{-1}$ (dotted curve: right axis) as functions of
 the redshifted mass $m_z$. The TDI method  was used here again.}
\label{f5}
\end{figure}

\end{document}